**High Field Magnet Development for HEP in Europe – A Proposal from LDG HFM Expert Panel**

P. Védrine (Chair)[a], L. García-Tabarés (Co-Chair)[b], B. Auchmann[c], A. Ballarino[d],

B. Baudouy[a], L. Bottura[d], P. Fazilleau[a], M. Noe[e], S. Prestemon[f], E. Rochepault[a], L. Rossi[g],

C. Senatore[h], B. Shepherd[i]

[a] CEA, Saclay, France

[b] CIEMAT, Madrid, Spain

[c] PSI, Villigen, Switzerland

[d] CERN, Geneva, Switzerland

[e] KIT/ITEP, Karlsruhe, Germany

[f] LBNL, Berkeley, California, USA

[g] University of Milan, Physics department and INFN division of Milan, LASA Laboratory, Milan, Italy

[h] University of Geneva, Switzerland

[i] STFC Rutherford Appleton Laboratory, Harwell Campus, UK

The European Laboratory Directors Group (LDG) was mandated by CERN Council in 2021 to oversee the development of an Accelerator R&D High Energy Physics Accelerator Roadmap. To this end, a set of expert panels was convened, covering the five broad areas of accelerator R&D highlighted in the ESPPU, drawing upon the international accelerator physics community for their membership, and tasked to consult widely and deeply.

High Field Magnets (HFM) is one of this R&D axes and are among the key technologies that will enable the search for new physics at the energy frontier. Approved projects (HL-LHC) and potential future circular machines such as proton-proton Future Circular Collider (FCC-hh) and Super proton-proton Collider (SppC) require the development of Superconducting (SC) magnets that produce fields beyond those attained in the LHC.

The present state of the art in HFM is based on $Nb_3Sn$, with magnets producing fields in the range of 11 T to 14 T. We have tackled in the last years the challenges associated with the brittle nature of this material, but we realize that more work is required and that manufacturing is not robust enough to be considered ready at an industrial scale.

Great interest has been also stirred in recent years by the progress achieved on HTS, not only in the fabrication of demonstrators for particle physics, but also in the successful test of magnets in other fields of application such as fusion and power generation. This shows that the performance of HTS magnets will exceed that of the $Nb_3Sn$, and also that the two technologies can be complementary to produce fields in the range of 20 T, and possibly higher.

The programme proposed here will advance beyond the results achieved over the past twenty years in past European and international programmes such as EU FP6 Coordinated Accelerator Research in Europe (CARE), EU FP7 European Coordination for Accelerator Research & Development (EuCARD), EU FP7 Enhanced European Coordination for Accelerator Research & Development (EuCARD2), EU FP7 Accelerator Research and Innovation for European Science and Society (ARIES), and current work such as HL-LHC, EU H2020 Innovation Fostering in Accelerator Science and Technology (I-FAST), CERN-HFM and US Magnet Development Program (US-MDP).

The proposed R&D programme (see [1] for a full description) has two main objectives. The first is to demonstrate Nb3Sn magnet technology for large-scale deployment. This will involve pushing it to its practical limits in terms of ultimate performance (towards the 16 T target required by FCC-hh), and moving towards production scale through robust design, industrial manufacturing processes and cost reduction, taking as a reference the HL-LHC magnets, i.e. 12 T). The second objective is to demonstrate the suitability of High Temperature Superconductor (HTS) for accelerator magnet applications, providing a proof-of-principle of HTS magnet technology beyond the range of Nb3Sn, with a target in excess of 20 T. The above goals are indicative, since the decision on a cost-effective and practical operating field will be one of the main outcomes of the development work.

The proposed roadmap (Fig. 1) comprises three focus areas (Nb3Sn and HTS conductors, Nb3Sn magnets, and HTS magnets) enabled by three cross-cutting activities (materials, cryogenics and models, powering and protection, and infrastructure and instruments).

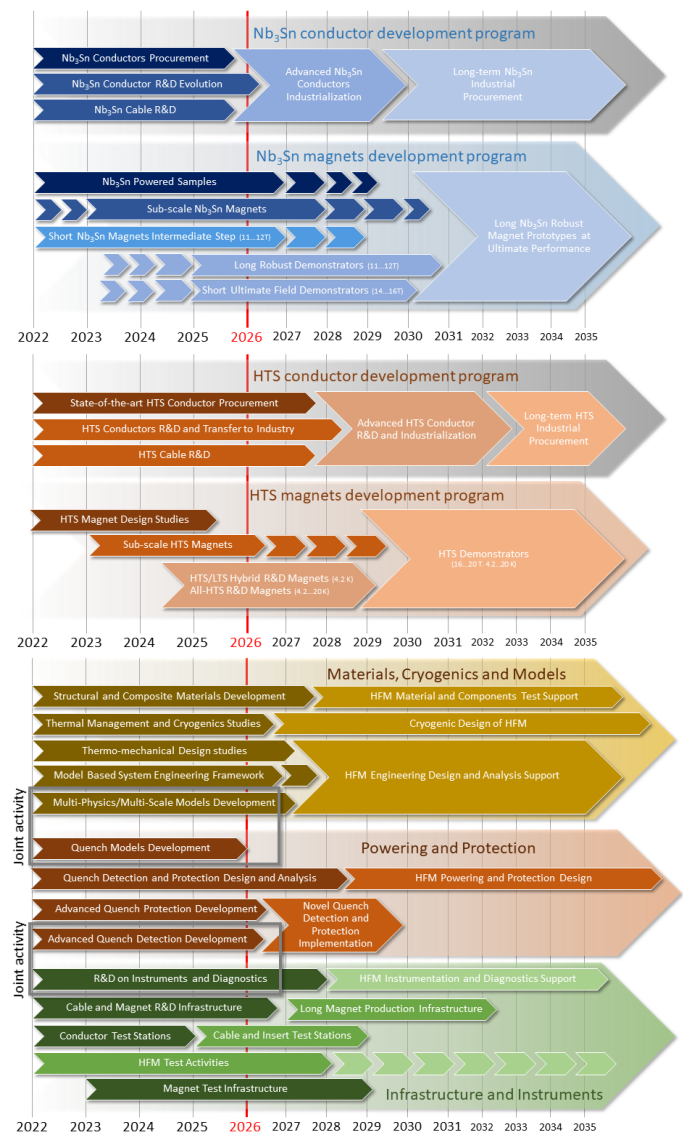

Fig. 1: Overview of proposed roadmap for high-field magnet development and associated technologies.

The methodology of the proposed programme is based on sequential development happening in steps of increasing complexity and integration, from samples, to small scale magnets, short magnets and long magnets in order to produce a fast-moving technology progression. We are convinced that fast-tracking and innovation are crucial to meeting the declared goals on a reasonable time scale.

The conductor activities, besides the necessary procurements, will focus on two aspects. $Nb_3Sn$ R&D will push beyond the state-of-the-art to consolidate the critical current capability (target non-copper current density of 1500 A/mm2 at 16 T and 4.2 K), establishing robust wire and cable configurations with reduced cost. These will then be the subject of a four-year period of industrialisation, which will be followed by a similar period of industrial optimisation. On the HTS side, the intention is to identify and qualify suitable tapes and cables, and follow up with industrial production to ensure the feasibility of large unit lengths (target 1 km) of HTS tapes with characteristics tailored to accelerator magnet applications. This HTS conductor R&D phase is expected to last for seven years.

The $Nb_3Sn$ magnet development will improve areas of HL-LHC technology that have been found to be sub-optimal, notably the degradation associated with the fragile conductor, targeting the highest practical operating field that can be achieved. The plan is to work jointly with wire and cable development to mitigate degradation associated either with length or electro-thermo-mechanical effects. The R&D will explore design and technology variants to identify robust design options for the field level targeted.

For $Nb_3Sn$ magnets, two objectives have been defined: the development of a 12T demonstrator of proven robustness suitable for industrialization, in parallel to the development of an accelerator demonstrator dipole reaching the ultimate field for this material, towards the target of 16 T.

The magnet technology R&D will progress in steps over a projected period of seven years, but is intended to provide crucial results through demonstration magnets in time for the next update of the European Strategy for Particle Physics (ESPP). Another five years are expected to be necessary to extrapolate the demonstrator results to full-length units.

R&D plans for HTS magnets focus on manufacturing and testing of sub-scale and insert coils as a vehicle to demonstrate performance and operation beyond the range of $Nb_3Sn$. For HTS magnets, a dual objective is proposed: the development of a hybrid LTS/HTS accelerator magnet demonstrator and a full HTS accelerator magnet demonstrator, with a target of 20 T. Special attention will be devoted to the possibility of operating in an intermediate temperature range (10K to 20 K).

The projected duration of this phase of test magnets, i.e. not yet accelerator designs, is seven years. By this time the potential of HTS for accelerator operation will be clear. At least five more years will be required to develop HTS demonstrators that include all the necessary accelerator features, surpassing $Nb_3Sn$ performance or working at temperatures higher than liquid helium. $Nb_3Sn$ is today the natural reference for future accelerator magnets, but HTS represents a real opportunity provided the current trend of production and price reduction is sustainable.

Energy efficiency efforts in line with societal trends should also be retained as one of the objectives when developing the next generation of magnets. The use of HTS conductors operated at higher temperatures could be a step in the right direction.

The cross-cutting technology activities will be a key seed for innovation. The scope includes materials and composites development using advanced analytics and diagnostics, new engineering solutions for the thermal management of high-field magnets, and the development of modelling tools within a unified engineering design framework. We propose to explore alternative methods of detection and protection against quench (especially important for HTS) including new measurement methods and diagnostics.

Finally, dedicated manufacturing and test infrastructure required for the HFM R&D programme, including instrumentation upgrades, needs to be developed, built and operated through close coordination between the participating laboratories.

The distribution of effort between the different axes of the nominal scenario described in the HFM programme is presented in Figure 2. The Nb3Sn conductor activities require a significant investment in the procurement of superconductor, about 50% of the total material value of the activities on Nb3Sn conductor. This procurement only marginally contributes to the conductor R&D, but is obviously necessary to feed the magnet development. The case is different for the HTS conductor, where tape and cable R&D dominates the cost of the programme.

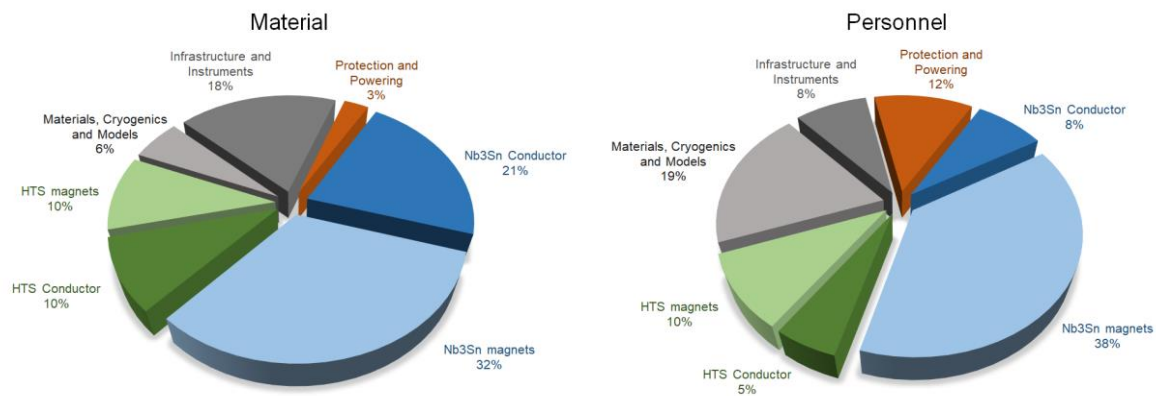

Fig. 2: Distribution of the effort between the different axes of the nominal scenario of the HFM programme.

The proposed programme also highlights to some extent the impact of the development of HFM magnets on the industrial ecosystem and on the training and education of future generations of applied scientists. One of the objectives is to propose actions to support European industry, responding to the ongoing evolution of business models, and fostering the deployment of developments and innovations from research to industry.

The programme would also like to emphasise the values of collaboration, and the connection to the ongoing programmes worldwide. Realizing the proposed HFM programme will build a broad and resilient basis of competence, a strong community, and the opportunity to educate the future generation on subjects of high-technological content.